\documentclass[journal]{IEEEtran}

\usepackage[pdftex]{graphicx}
\usepackage{booktabs}
\usepackage[colorlinks=true, linkcolor=black, citecolor=black, urlcolor=black]{hyperref}
\usepackage{tabularx}
\usepackage{cite}

\begin{document}
\title{Deep models for stroke segmentation: do complex architectures always perform better?}

\author{ Yalda Zafari-Ghadim$^{a}$ Ahmed Soliman$^{a}$, Yousif Yousif$^{a}$,  Ahmed Ibrahim$^{a}$, Essam A. Rashed$^{b}$ and Mohamed Mabrok$^{a}$

\thanks{ $^{a}$ Mathematics Program, Department of Mathematics, Statistics and Physics, College of Arts and Sciences, Qatar University, P.O.Box 2713, Doha, Qatar.}
 \thanks{$^{b}$Graduate School of Information Science, University of Hyogo, Kobe 650-0047, Japan}
}

\maketitle

\begin{abstract}
Stroke segmentation plays a crucial role in the diagnosis and treatment of stroke patients by providing spatial information about affected brain regions and the extent of damage. Segmenting stroke lesions accurately is a challenging task, given that conventional manual techniques are time-consuming and prone to errors. Recently, advanced deep models have been introduced for general medical image segmentation, demonstrating promising results that surpass many state-of-the-art networks when evaluated on specific datasets. With the advent of the vision Transformers, several models have been introduced based on them, while others have aimed to design better modules based on traditional convolutional layers to extract long-range dependencies like Transformers. The question of whether such high-level designs are necessary for all segmentation cases to achieve the best results remains unanswered. In this study, we selected four types of deep models that were recently proposed and evaluated their performance for stroke segmentation: (1) a pure Transformer-based architecture (DAE-Former), (2) two advanced CNN-based models (LKA and DLKA) with attention mechanisms in their design, (3) an advanced hybrid model that incorporates CNNs with Transformers (FCT), and (4) the well-known self-adaptive nnU-Net framework with its configuration based on given data. We examined their performance on two publicly available datasets, ISLES 2022 and ATLAS v2.0, and found that the nnU-Net achieved the best results with the simplest design among all. Additionally, we investigated the impact of an imbalanced distribution of the number of unconnected components in each slice, as a representation of common variabilities in stroke segmentation. Revealing the robustness issue of Transformers to such variabilities serves as a potential reason for their weaker performance. Furthermore, nnU-Net's success underscores the significant impact of preprocessing and postprocessing techniques in enhancing segmentation results, surpassing the focus solely on architectural designs.

\end{abstract}


\IEEEpeerreviewmaketitle

\section{Introduction}

Stroke, the second leading cause of morbidity and mortality worldwide, occurs due to sudden disruptions in cerebral blood flow that result in neurocellular damage or death \cite{feigin2022world, meyer2015systematic}. Medical imaging modalities such as magnetic resonance imaging (MRI) and computed tomography (CT) offer valuable information on stroke location, time, and severity \cite{goldstein2005patient, hwang2012comparative, simonsen2015sensitivity}. Stroke segmentation plays a crucial role by providing spatial information about affected brain regions and the extent of damage, aiding in diagnosis and treatment. However, accurate segmentation of stroke lesions remains challenging, as traditional manual methods are prone to errors and time consuming.

Deep neural networks have achieved state-of-the-art results in numerous computer vision tasks, including medical image segmentation, by learning intrinsic patterns in a data-driven manner \cite{zhou2019handbook}. Convolutional neural networks (CNNs) have been extensively utilized in segmentation architectures like U-Net \cite{ronneberger2015unet}. However, Conv-based models generally have limitations in modeling long-range dependencies and providing global information due to their limited receptive field \cite{chen2021transunet}, which is essential for tasks that require a global understanding of the entire input. Vision Transformers \cite{dosovitskiy2020image, liu2021swin}, inspired by the Transformer model introduced in natural language processing \cite{vaswani2017attention}, can model long-range dependencies in data, allowing the network to capture complex spatial relationships. The core idea in Transformers is the self-attention mechanism \cite{vaswani2017attention}, which enables models to weigh different parts of the input differently and consider the most relevant parts. In general, vision Transformers lack some useful inductive biases that Conv-based models have, such as translation equivariance and locality. To address this issue and capture more comprehensive information from the data, hybrid Conv-Transformer models have been widely used \cite{hatamizadeh2021swinunetr, hatamizadeh2022unetr, zhou2021nnformer, zhang2021transfuse}.

Most previous research on stroke segmentation using deep models is based on Conv-based U-shaped architectures \cite{clerigues2019acute, abbasi2023automatic, khezrpour2022automatic}, but in recent years some hybrid Conv-Transformer models have also been introduced \cite{marcus2023concurrent, gu2022sthardnet}. \cite{zhou2019d} incorporated a fusion of 2D and 3D convolutions to facilitate an effective representation of infarcts to enhance the segmentation results. Some research studies employed additional blocks or advanced versions of convolutions to increase the receptive field, to extract local and global information. For instance, \cite{abulnaga2019ischemic} employed the pyramid pooling module \cite{zhao2017pyramid}, where various kernel sizes were used to aggregate multi-scale features from the data. Another study by \cite{liu2019efficient} also employed multi-kernels of various sizes to extract feature information across different receptive fields. Pool-UNet, introduced by \cite{liu2022pool}, employed Squeeze and Excitation blocks \cite{hu2018squeeze} in the bottleneck to provide more informative features for the decoder path by capturing interdependencies between channels. Additionally, a combination of CNNs and the Poolformer structure \cite{yu2022metaformer} was utilized to benefit from both local and global information. \cite{chalcroft2023large} employed Large Kernel Attention \cite{guo2023visual} to take advantage of the inherent biases of CNNs while capturing long-range dependencies. 

Some studies have introduced hybrid models to leverage the strengths of CNNs and Transformers for more accurate stroke segmentation. In UCATR \cite{luo2021ucatr}, an architecture was proposed consisting of a CNN-based encoder and decoder, with a Transformer module facilitating the extraction of global information. By using multi-head cross-attention modules in the skip connections, irrelevant information was filtered in this model. In another work, a model named UTransNet \cite{feng2022utransnet} was introduced, where the main blocks of the encoder and decoder paths consisted of two convolutional layers followed by a Transformer module. In the METrans model \cite{wang2022metrans}, multiple Convolutional Block Attention Modules (CBAM) \cite{woo2018cbam} were used to take advantage of channel and spatial attention. Incorporating a Transformer block in the bottleneck of the U-shaped architecture proved useful for extracting global information. A dual-path model was also proposed, known as LLRHNet \cite{liu2022llrhnet}, which consisted of a CNN encoder and a Transformer-based encoder; the extracted features from these encoders were then fused. Some studies also employed hybrid models to mitigate the issue of oversmooth boundary estimation for fuzzy stroke infarcts in their proposed models \cite{wu2023transrender, wu2023w}.

Although variabilities between patients, vendors, imaging artifacts, and noise present challenges for deep networks in most medical image segmentation tasks, this issue is more severe for stroke segmentation. Unlike organ segmentation tasks, such as cardiac segmentation, where certain characteristics like shape and location may be common across cases, stroke segmentation lacks such universally applicable features. Variations in stroke infarct size, number, location, shape, pattern, and difference in intensity based on their age pose significant difficulties for deep models. These unresolved challenges make the current proposed pipelines less applicable to accurately segmenting the stroke \cite{zhang2022application}.

Recently, advanced deep models have been introduced for general medical image segmentation, showcasing promising results that surpass many state-of-the-art networks when evaluated on specific datasets. However, these models have not been evaluated for stroke segmentation. In this paper, we selected four architectures that were recently proposed and evaluated their performance for stroke segmentation using two publicly available datasets. To gain a better understanding of models based on their design by CNNs or Transformers for stroke segmentation, we included a pure Transformer-based model (DAE-Former), two CNN-based models (LKA and DLKA), an advanced model that incorporates CNNs within Transformers (FCT), and the self-adaptive framework of nnU-Net.

The paper's contribution can be summarized as follows: 
\begin{itemize}
    \item The paper evaluates the performance of four different types of recently developed deep models for medical image segmentation for stroke segmentation.

\item The evaluation is conducted on two large datasets with different modalities, ISLES 2022 and ATLAS v2.0, providing a comprehensive comparison across different architectures.
\item The study identifies that the nnU-Net framework achieves the best results among all evaluated models for stroke segmentation, despite its simpler design compared to Transformer-based architectures and other advanced CNN-based models.
\item The paper sheds light on the importance of convolutional layers within stroke segmentation architectures, highlighting their ability to capture local information, which is crucial for accurate segmentation. Also, the paper underscores the significant impact of preprocessing and post-processing techniques in enhancing segmentation results, suggesting that focusing solely on architectural designs may not be sufficient for achieving optimal performance in medical image segmentation tasks like stroke segmentation.
\end{itemize}

The paper is organized as follows: Section 2 provides an overview of the deep architectures employed in this study for stroke segmentation, along with the utilized datasets. Section 3 provides information on the evaluation metrics used to assess the performance of different architectures, implementation details, and the training process. It also summarizes our findings and includes a comparison of the performance of different models on two public datasets with a discussion related to various aspects of stroke segmentation using recently introduced deep architectures. Finally, Section 4 presents our conclusions based on the results of this research effort and highlights the limitations of our experiments.

\section{Architectures and Datasets}
In this section, we briefly survey four different types of recently developed deep models for medical image segmentation, which will be used in our evaluation for stroke segmentation. The architectures were selected based on their architectural designs and superior performance compared to other networks. Table \ref{tab:synapse} summarizes their performance for different benchmark datasets including  Synapse\footnote{\url{https://doi.org/10.7303/syn3193805}}, ACDC \cite{bernard2018deep}, and  ISIC 2018 \cite{codella2019skin}, and the state-of-the-art methods they surpassed, as mentioned in their original papers.

\begin{table*}
  \centering
  \caption{Comparison of performance across different architectures on different datasets, including the baseline networks they have surpassed in terms of performance as reported in the original paper.}
    \begin{tabularx}{\textwidth}{|l|c|c|c|X|}
    \hline
    \textbf{Architectures} & \textbf{Synapse} & \textbf{ACDC} & \textbf{ISIC 2018} & \textbf{Surpassed Networks}  \\
    \hline
    DAE-Former \cite{azad2023dae} & 0.8263 & -  & 0.9147 & U-Net \cite{ronneberger2015unet}, Att-UNet \cite{oktay2018attention}, TransUNet \cite{chen2021transunet}, Swin-UNet \cite{cao2022swin}, LeViT-UNet \cite{xu2023levit}, MT-UNet \cite{wang2022mixed}, TransDeepLab \cite{azad2022transdeeplab}, HiFormer \cite{heidari2023hiformer}, MISSFormer \cite{huang2021missformer}, MCGU-Net \cite{asadi2020multi}, MedT \cite{valanarasu2021medical}, FAT-Net \cite{wu2022fat}, TMU-Net \cite{azad2022contextual}, TransNorm \cite{azad2022transnorm}\\
    \hline
    FCT \cite{tragakis2023fully}& 0.8353 & 0.9302 & - & U-Net \cite{ronneberger2015unet}, Att-UNet \cite{oktay2018attention}, TransUNet \cite{chen2021transunet}, TransClaw U-Net \cite{chang2021transclaw}, Swin-UNet \cite{cao2022swin}, LeViT-UNet \cite{xu2023levit}, MT-UNet \cite{wang2022mixed}, nnFormer \cite{zhou2021nnformer}  \\ 
    \hline
    LKA \cite{azad2024beyond}& 0.8277 & - & 0.9118 &  TransUNet \cite{chen2021transunet}, Swin-UNet \cite{cao2022swin}, LeViT-UNet \cite{xu2023levit}, MISSFormer \cite{huang2021missformer}, Hi-Former \cite{heidari2023hiformer}, DAE-Former \cite{azad2023dae}, TransDeepLab \cite{azad2022transdeeplab}, PVT-CASCADE \cite{rahman2023medical}, DAGAN \cite{lei2020skin}, MCGU-Net \cite{asadi2020multi}, MedT \cite{valanarasu2021medical}, FAT-Net \cite{wu2022fat}, TMU-Net \cite{azad2022contextual}, DeepLabv3+ \cite{chen2018encoder}  \\ 
    \hline
    D-LKA \cite{azad2024beyond}& 0.84277 & - & 0.9177 & TransUNet \cite{chen2021transunet}, Swin-UNet \cite{cao2022swin}, LeViT-UNet \cite{xu2023levit}, MISSFormer \cite{huang2021missformer}, ScaleFormer \cite{huang2022scaleformer}, Hi-Former \cite{heidari2023hiformer}, DAE-Former \cite{azad2023dae}, TransDeepLab \cite{azad2022transdeeplab}, PVT-CASCADE \cite{rahman2023medical}, DAGAN \cite{lei2020skin}, MCGU-Net \cite{asadi2020multi}, MedT \cite{valanarasu2021medical}, FAT-Net \cite{wu2022fat}, TMU-Net \cite{azad2022contextual}, DeepLabv3+ \cite{chen2018encoder}  \\
    \hline
    nnU-Net \cite{isensee2021nnu} & 0.8880 \cite{tang2022self}& 0.9161 \cite{tragakis2023fully} & - & -   \\ 
    \hline
  \end{tabularx}
  \label{tab:synapse}
\end{table*}

\subsection{Architectures}

\subsubsection{nnU-Net}
\begin{figure}
  \centering
  \includegraphics[width=0.48\textwidth]{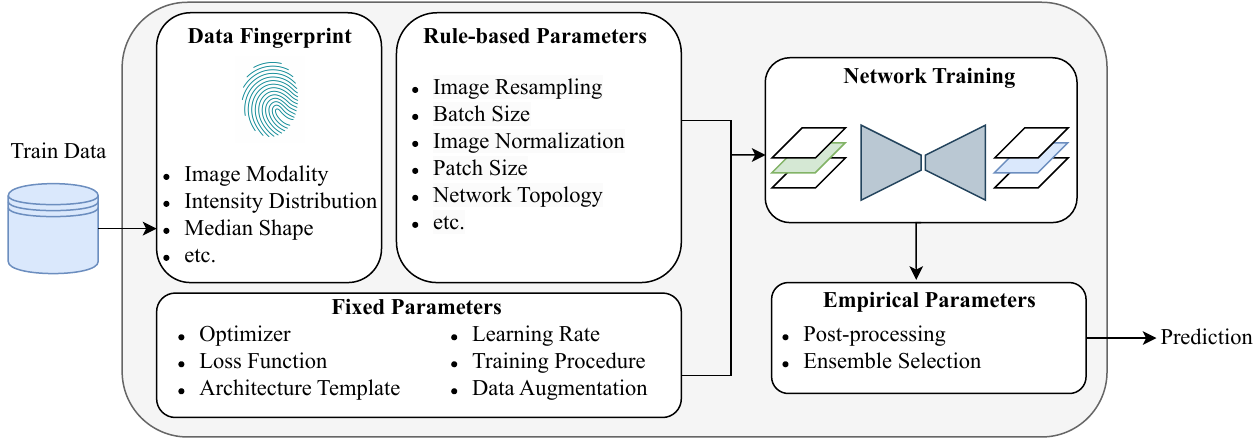}
  \caption{nnU-Net's pipeline for designing a segmentation network for each dataset \cite{isensee2021nnu}. }
  \label{fig:nnunet}
\end{figure}

The no new U-Net (nnU-Net) framework stands out as an exceptional tool, building on the well-established U-Net architecture and incorporating innovative self-adaptive capabilities \cite{isensee2021nnu}. In contrast to traditional methods that rely on manual tuning, nnU-Net autonomously adjusts itself for new tasks, simplifying key aspects from preprocessing to post-processing with minimal human involvement. What distinguishes nnU-Net is its straightforwardness and emphasis on fundamental elements, resulting in remarkable performance across various challenges. It has demonstrated its effectiveness by surpassing existing methods across 23 public datasets, highlighting its adaptability and resilience in diverse segmentation tasks, including stroke segmentation \cite{el2022evaluating, huo2022mapping}. Departing from the original U-Net architecture, nnU-Net incorporates leaky ReLUs and instance normalization, underscoring the potential efficacy of non-architectural adjustments over architectural modifications. See Figure \ref{fig:nnunet} for an illustration of nnU-Net's pipeline for automated configuration when provided with a new dataset.

\subsubsection{DAE-Former}
\begin{figure}
  \centering
  \includegraphics[width=0.4\textwidth]{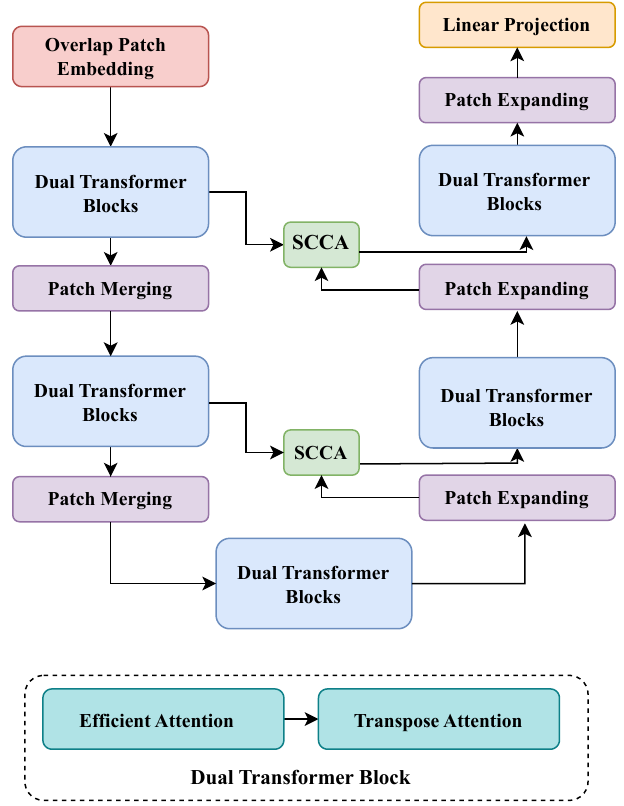}
  \caption{Structure of the DAE-Former architecture \cite{azad2023dae}. The central component of this architecture is the Dual Transformer, comprising two distinct attention mechanisms: efficient attention for capturing spatial information and transpose attention for capturing channel information. Additionally, the skip connection cross-attention (SCCA) module is employed to integrate information from encoder layers with features from decoder layers. This fusion process enhances extracted features by preserving the most relevant information. }
  \label{fig:daeformer}
\end{figure}

Dual Attention-guided Efficient Transformer (DAE-Former) is a pure Transformer-based U-shape architecture designed for medical image segmentation \cite{azad2023dae}. It has demonstrated superior performance over many state-of-the-art architectures for multi-organ cardiac and skin lesion segmentation tasks. This architecture incorporates three types of attention mechanisms: (1) efficient attention \cite{shen2021efficient}, (2) transpose attention \cite{ali2021xcit}, and (3) cross attention. See Figure \ref{fig:daeformer} for an illustration of the DAE-Former structure.

By incorporating efficient attention and transpose attention, dual-attention blocks were developed and utilized in both the encoder and decoder paths. These blocks aim to capture both spatial and channel information, thereby enhancing the model's capacity to capture more informative feature information. The dual attention blocks start with patch embedding modules, extracting overlapping patch tokens from the input image. Efficient attention is a modified version of the self-attention mechanism, the core element of Transformers. It addresses the quadratic computational complexity issue, making it more applicable for use with high-resolution medical images. In regular Transformers, after producing the query (Q), key (K), and value (V) vectors from the input sequence by mapping it onto the embedding dimension (d), self-attention is calculated as follows:
\begin{equation}
   SA(Q, K, V) = Softmax(\frac{Q{\times}K^T}{\sqrt{D}}){\times}V.
\end{equation}

Efficient attention utilizes normalization functions ($\rho$) for the query and key, employing \emph{softmax} normalization functions, and then calculates the multiplications to produce the final representation. The process can be formulated as:

\begin{equation}
   EA(Q, K, V) = {\rho}_q(Q){\times}({\rho_k(K)}^T{\times}V).
\end{equation}
For capturing channel-related information, cross-covariance attention, also known as transpose attention, is employed. This transpose attention is calculated as follows:

 \begin{equation}
   TA(Q, K, V) = V{\times}Softmax(\frac{K^T{\times}Q}{\tau}),
\end{equation}
where $\tau$ is employed to restrict the magnitude of the query and key matrices through $\ell_2$-normalization, improving stability during training. Another crucial attention module is skip-connection cross-attention (SCCA). Skip connections, commonly utilized in U-shaped architectures, help preserve the flow of information across different scales. While plain skip connections typically involve concatenating features extracted from encoder layers to decoder layers, advanced versions such as attention-based connections can be employed to suppress irrelevant regions in the image during training and highlight significant features for the segmentation task. In DAE-Former, the efficient attention structure is utilized for SCCA, with the distinction that the input used for the query originates from the encoder layers, while the inputs for keys and values are derived from the decoder layers.

\subsubsection{FCT}
This architecture, introduced as the first fully convolutional Transformer (FCT), is designed to be a robust and accurate network for medical image segmentation \cite{tragakis2023fully}. It incorporates CNN layers into the architecture of Transformers to leverage CNNs' ability to represent image features and Transformers' capability to model long-range dependencies. In FCT, all projection layers in the Transformer architecture are replaced by depth-wise convolutional layers, providing an option to reduce computational costs and obtain a better spatial context from images.

\begin{figure*}
  \centering
  \includegraphics[width=0.9\textwidth]{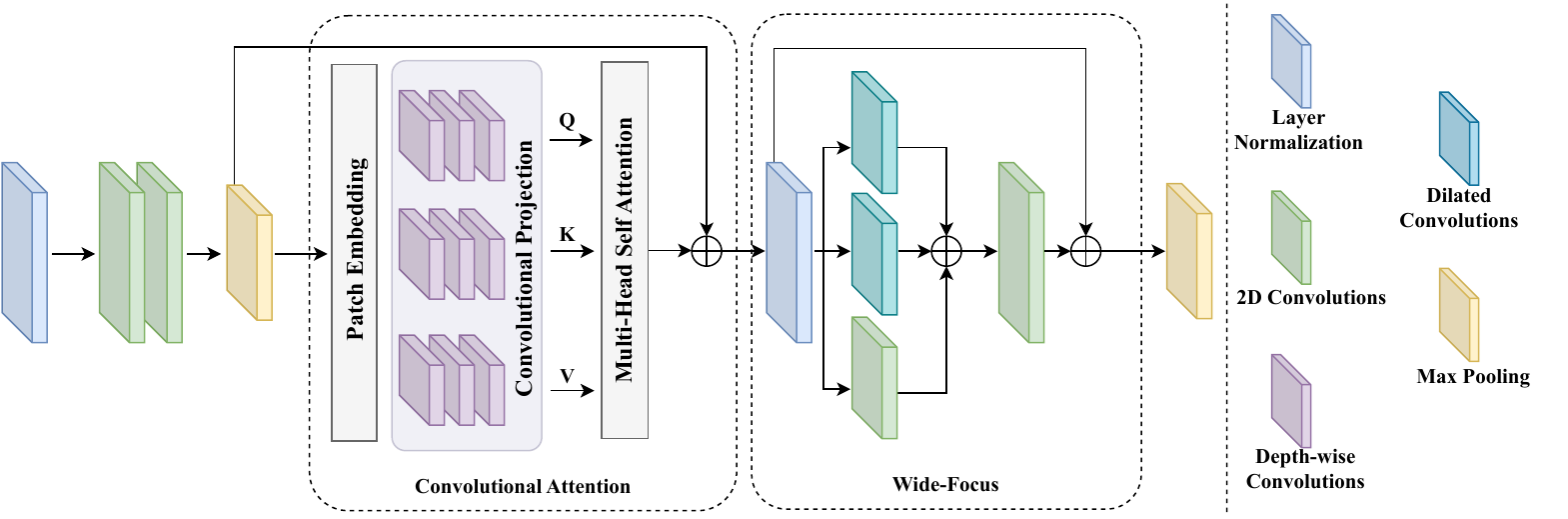}
  \caption{Illustration of the main blocks in FCT \cite{tragakis2023fully}. Each block comprises two key modules: Convolutional Attention and Wide Focus. Following initial convolutional layers, data undergoes processing through Convolutional Attention, akin to conventional vision Transformers. However, instead of MLP layers serving as projection layers, three layers of depth-wise convolutions are employed. Subsequently, the output is fed into the Wide Focus module, which consists of three convolutional layers: one standard and two dilated layers with varying kernel sizes to augment the receptive field. The outputs of these layers are summed and passed through an additional convolutional layer for further processing.  }
  \label{fig:fct}
\end{figure*}

The resulting overlapping patches are fed into the Transformer encoder block to apply the self-attention mechanism. Each FCT layer starts with a sequence of operations: Layer Normalization, Convolution, Convolution, and MaxPool. These convolutional layers are followed by a GELU activation function. The output of the Transformer encoder block is then processed through the Wide-Focus module, which is a multi-branch convolutional paradigm adapted to enhance spatial context. One layer applies spatial convolution, while the others apply dilated convolutions with increasing receptive fields. Extracted features from multiple branches are fused via summation and passed into a spatial convolution operator as a feature aggregation layer. See Figure \ref{fig:fct} for an illustration of the main blocks utilized in FCT architecture.

Additionally, the architecture adopts a multi-scale input for its U-shaped design by concatenating resampled versions of the input to the encoder features at multiple scales. This approach enhances the model's ability to capture features at different resolutions, contributing to improved segmentation performance.

\subsubsection{D-LKA}
LKA (Large Kernel Attention) employs convolutional kernels significantly larger than those in traditional methods, capturing a broader contextual view akin to the receptive fields observed in self-attention mechanisms \cite{guo2023visual}. Remarkably, LKA achieves this expanded field of view with substantially fewer parameters compared to self-attention, leading to improved computational efficiency utilizing a strategic combination of depth-wise and dilated convolutions within these large kernels. The utility of LKA in medical image segmentation, particularly in improving the representation of complex structures such as lesions or organ deformations, is underscored by its integration with Deformable Convolutions, further refining the method's adaptability and accuracy \cite{dai2017deformable}.

\begin{figure}
  \centering
  \includegraphics[width=0.4\textwidth]{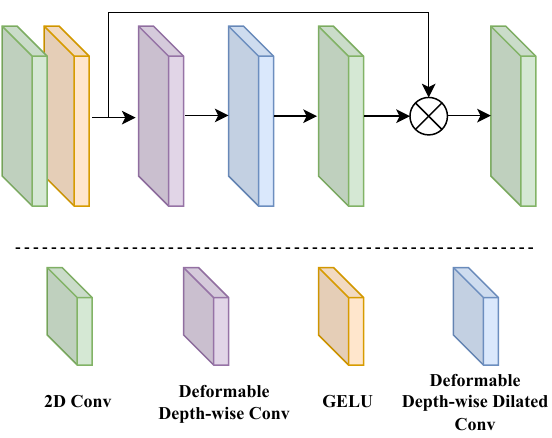}
  \caption{Illustration of the main blocks in D-LKA \cite{azad2024beyond}. Following processing by a convolutional layer, the data is inputted into the large kernel attention block. This block first applies a sequence of operations: a deformable depth-wise convolutional layer, followed by a deformable depth-wise dilated convolutional layer, and finally a regular convolutional layer. The attention mechanism, implemented through multiplication, serves to suppress irrelevant information, facilitating effective learning during training. Notably, in the LKA architecture, all convolutions are non-deformable, while maintaining the same underlying structure. }
  \label{fig:dlka}
\end{figure}

\begin{figure}
  \centering
  \includegraphics[width=0.5\textwidth]{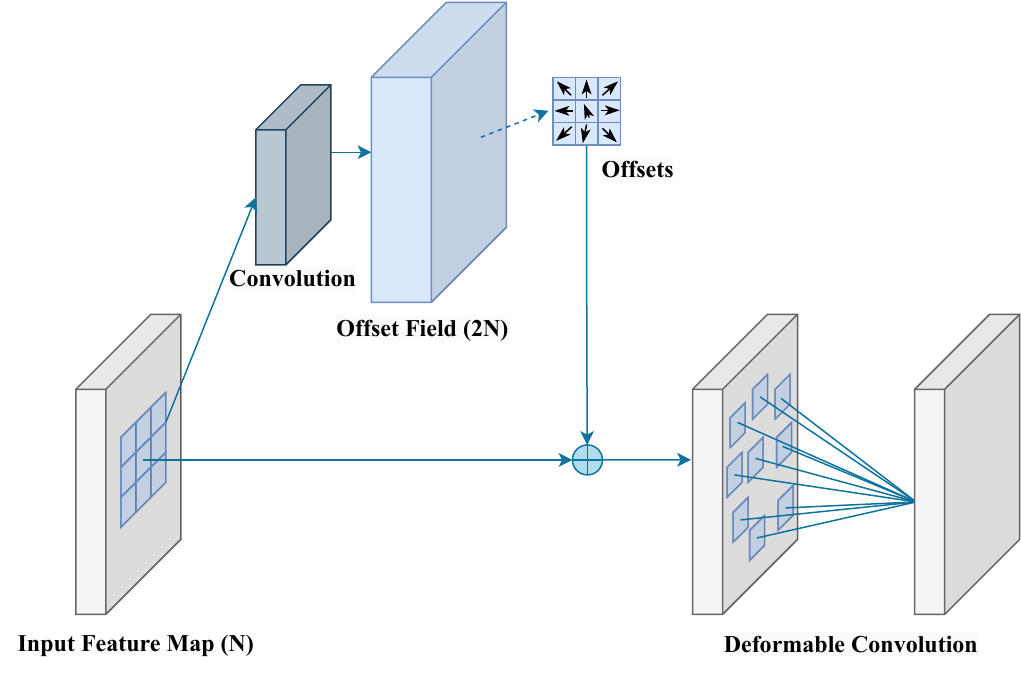}
  \caption{Illustration depicting the concept of deformable convolutions, a technique that enhances standard grid sampling positions used in regular convolutions by integrating 2D offsets. This modification enables the sampling grid to flexibly deform, with the offsets learned from preceding feature maps through additional convolutional layers. As a result, this approach conditions deformation based on input features, offering localized, dense, and adaptable adjustments. }
  \label{fig:deformable}
\end{figure}

Unlike LKA, which uses fixed kernel sizes and shapes, Deformable Large Kernel Attention (DLKA) introduces the ability to dynamically adjust the sampling grid \cite{azad2024beyond}. This adjustment is achieved by applying whole-numbered offsets for free deformation, learned from the feature maps themselves. Refer to Figure \ref{fig:dlka} for a visualization of the main blocks employed in DLKA and LKA architecture, while Figure \ref{fig:deformable} demonstrates the functionality of deformable convolutions. Consequently, DLKA facilitates an adaptive convolution kernel capable of flexibly adjusting its shape to better represent complex anatomical structures. DLKA employs an additional convolutional layer dedicated to learning these deformations and creating an offset field. This layer aligns with the kernel size and dilation rate of its corresponding convolutional layer in the DLKA module. For this study, we evaluated both D-LKA and LKA architectures to assess the impact of deformable-based convolutions compared to regular ones.

\subsection{Datasets}
Among the datasets available for stroke segmentation, we selected two based on their accessibility and the number of data they offer. This decision ensures a sufficiently large dataset for training deep neural networks effectively.

\subsubsection{ISLES 2022}
The Ischemic Stroke Lesion Segmentation (ISLES) dataset serves as an important resource in the field of stroke lesion segmentation. The 2022 version of ISLES comprises 400 MRI cases sourced from multiple vendors, with 250 publicly accessible cases and 150 private ones \cite{hernandez2022isles}. This dataset offers a comprehensive view of ischemic stroke lesions, showcasing diverse infarct patterns, variable lesion sizes, and locations. Furthermore, the heterogeneity of the data set, resulting from the use of imaging devices from three different medical centers, presents a valuable opportunity to assess the generalization of the proposed methods. Modalities encompass DWI, ADC, and FLAIR images, making it a valuable resource for advancing research on stroke lesion segmentation. See Figure \ref{fig:isles} for some samples from this dataset.

\subsubsection{ATLAS v2.0}
The Anatomical Tracings of Lesions After Stroke (ATLAS) datasets \cite{liew2022atlas} are available in two versions: 1.2 and 2.0, both featuring high-resolution T1-weighted MRI images accompanied by the corresponding lesion masks. Version 1 comprises a total of 304 cases, whereas version \emph{2.0} is more extensive, containing 955 cases. Among these 955 cases, 655 T1-weighted MRI scans are publicly accessible along with their ground truth, while an additional 300 T1-weighted MRI scans lack corresponding lesion masks. Preprocessing steps applied to this dataset include intensity normalization and registration onto the MNI-152 template. Sample images from this dataset are shown later in Figure \ref{fig:atlas}.

\section{Results and Discussion}
In this section, we present the implementation details and performance assessment metrics.  Additionally, we discuss the performance of the networks on the two datasets.
\subsection{Implementation details}
The DAE-Former, LKA, and D-LKA architectures were implemented in PyTorch, while the FCT architecture was implemented using TensorFlow. Additionally, nnU-Net was trained using a self-adaptive method for 100 epochs, and most of the hyperparameters for training were chosen by the framework itself. All the results reported in this section were obtained using an NVIDIA A100 GPU with 80GB memory. The training utilized a batch size of 24 over 100 epochs with the Stochastic Gradient Descent (SGD) optimizer. A multi-step learning rate adjustment was used, starting at an initial rate of 0.05.

All the selected models were designed for 2D images, so we sliced the volumes of each patient in each dataset accordingly. For the ISLES2022 dataset, we utilized all planes - sagittal, coronal, and axial - to increase the volume of data, as it was lower compared to the ATLAS dataset. However, for the ATLAS dataset, we sliced the cases from the axial plane. All images were resampled to a size of $224 \times 224$.

During experiments on the ISLES 2022 dataset, we specifically utilized the DWI modality. This decision was made because the FLAIR modality had a different voxel spacing compared to DWI and ADC. Furthermore, we did not observe significant performance differences when using both DWI and ADC channels compared to using DWI alone. To enhance data diversity during training, we applied various data augmentation techniques, including random rotation, scaling, shear, translation, and flipping.

We experimented with various loss functions during the training of each architecture. The best results were achieved for LKA, D-LKA, and DAE-Former when trained with a fused loss function that combines the Dice loss and Binary Cross Entropy (BCE) loss, with weights of 0.6 and 0.4, respectively. However, the FCT architecture yielded the best results when trained solely on BCE loss. It's worth noting that D-LKA encountered challenges during training on the ATLAS v2.0 dataset. These difficulties may stem from the high number of parameters in this network \cite{azad2024beyond}, compounded by the dataset's severe imbalance issue, where approximately $82.17\%$ of all slices lacked stroke. Consequently, we excluded this network from the comparison on the ATLAS dataset.

\subsection{Performance Assessment}
\begin{figure*}
  \centering
  \includegraphics[width=\textwidth]{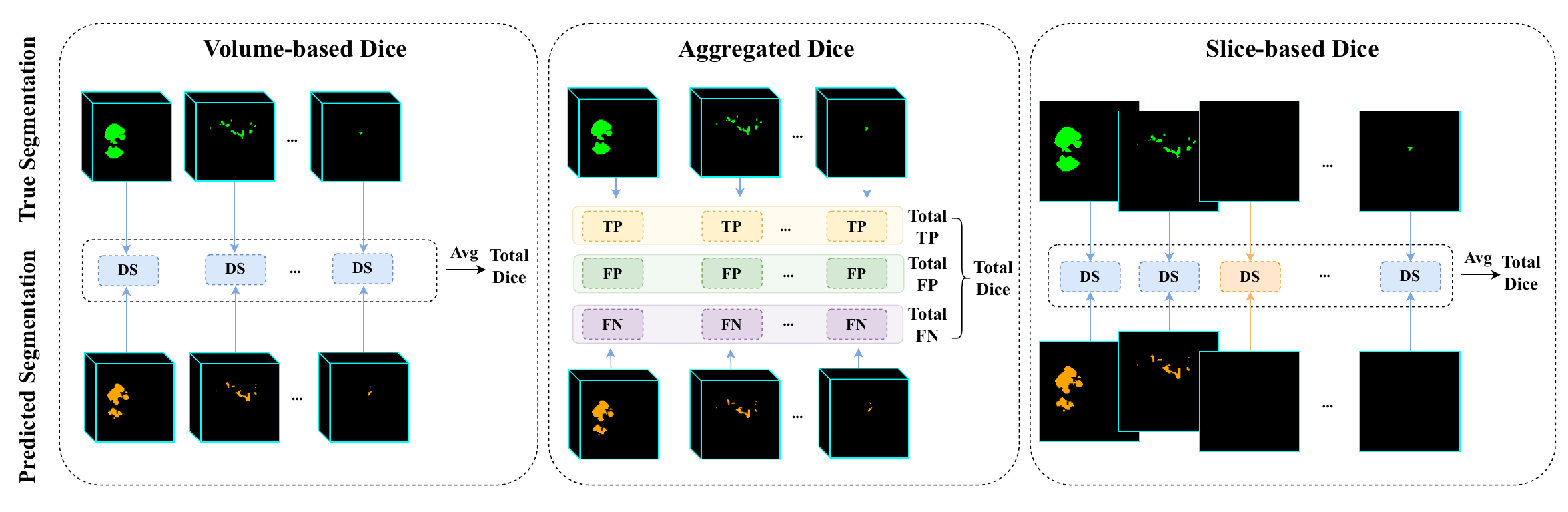}
  \caption{Different approaches for calculating the Dice score. In the volume-based approach, the Dice score is computed individually for each 3D volume, and the overall Dice score is obtained by averaging across all volumes. The aggregate approach involves tallying true positives (TP), false positives (FP), and false negatives (FN) for each slice or case, and the final Dice score is calculated by aggregating these values across all slices or cases. Meanwhile, the slice-based method computes the Dice score in a 2D fashion, where each slice is assessed independently. However, null slices (their Dice score is shown in orange), devoid of stroke in both ground truth and predicted labels, can significantly influence the results in this approach. }
  \label{fig:dice}
\end{figure*}

The Dice score (DS) serves as one of the primary evaluation metrics for segmentation tasks, particularly in imbalanced scenarios like stroke, providing a quantitative measure of segmentation accuracy. DS is a statistical metric used to measure the similarity between two sets by assessing the overlap between the predicted segmentation and the ground truth, yielding a value between 0 and 1, where 1 indicates perfect alignment, and 0 indicates no overlap. The DS between sets A and B is defined as:

\begin{equation}
    DS = \frac{2 |A \cap B|}{|A| + |B|} = \frac{2{\times}TP}{2{\times}TP + FP + FN}
\end{equation}

where TP, FP, and FN represent true positive, false positive, and false negative, respectively. One challenge encountered in utilizing DS arises from the variety of methods available for its calculation, complicating the comparison of results across different models and studies. To address this issue, three distinct approaches for calculating DS were employed here, each offering unique insights into the performance of the segmentation model. Refer to Figure \ref{fig:dice} for an illustration of the different ways of calculating the Dice score.

\subsubsection{Aggregated DS}
An aggregated way of calculating the Dice score can be utilized for both 2D slices and 3D volumes. This method aggregates the TP, FP, and FN values across all cases first. Following this aggregation, the DS is computed just once at the end of the process. Summing up the TP, FP, and FN values across all cases captures the overall performance of the segmentation process in the entire data set. This can be particularly beneficial in scenarios where a cumulative assessment is more informative than individual volume assessments. However, it's important to note that for stroke segmentation, where there are different sizes of stroke infarcts, this method tends to be biased towards the performance of segmentation for large infarcts, and the performance is typically better for larger infarcts compared to smaller ones.

\subsubsection{Slice-based DS}
The second method involves calculating the Dice score for each slice independently. This approach allows for a more detailed understanding of segmentation performance at a single-slice level. For stroke segmentation, where there can be different sizes in different slices, this method can offer better insight into performance. However, it is crucial to handle null slices carefully, those lacking stroke infarcts in both ground truth and predicted masks. Assigning a score of 1 to null slices could potentially inflate the average Dice score, especially in cases where only a limited number of slices are involved with stroke. This could result in a higher Dice score even with lower stroke segmentation performance. After calculating these individual scores, the mean of all Dice scores is computed, providing an average measure of segmentation accuracy throughout the entire dataset.

\subsubsection{Volume-based DS}
In the third method, a 3D volume-based calculation, the Dice score is computed separately for each volume. Subsequently, the mean of these scores is taken across all cases. This method provides a comprehensive assessment of segmentation performance by considering spatial relationships between voxels within each volume. However, it has a potential drawback. As stroke segmentation often involves multiple instances, there are cases with both large infarcts and several small infarcts in other regions simultaneously. This calculation method may neglect the small infarcts and yield a high Dice score even if the small infarcts were not segmented correctly.

\subsection{Results}
\begin{figure*}
  \centering
  \includegraphics[width=\textwidth]{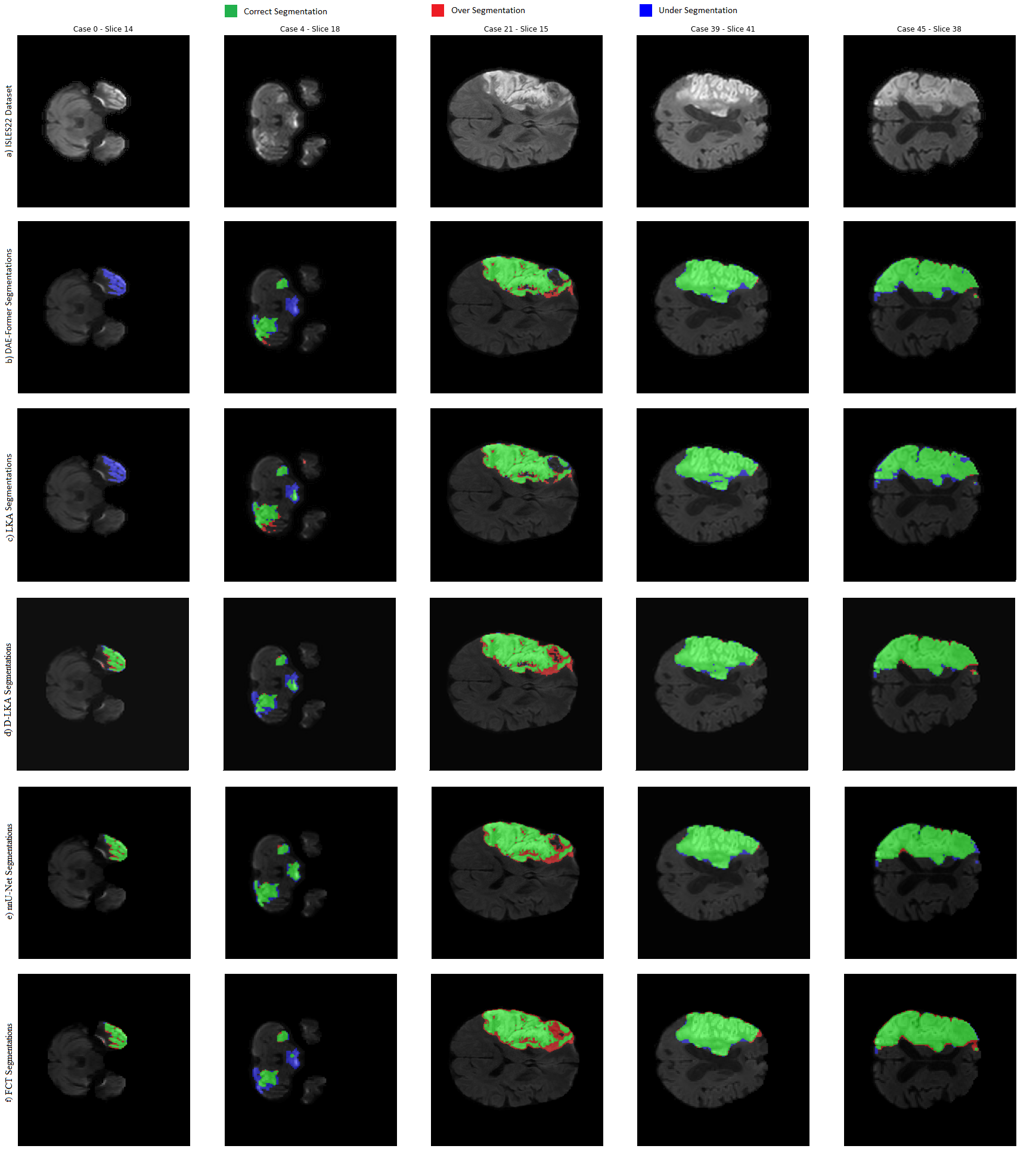}
  \caption{a) Sample of ISLES 2022 dataset,  b) Performance of the DAE-Former \cite{azad2023dae},   c)  Performance of the LKA \cite{azad2024beyond},  d)  Performance of the D-LKA, \cite{azad2024beyond},  e) Performance of the nnU-Net \cite{isensee2021nnu},  f)  Performance of the FCT \cite{tragakis2023fully}.}
  \label{fig:isles}
\end{figure*}

\begin{figure*}
  \centering
  \includegraphics[width=\textwidth]{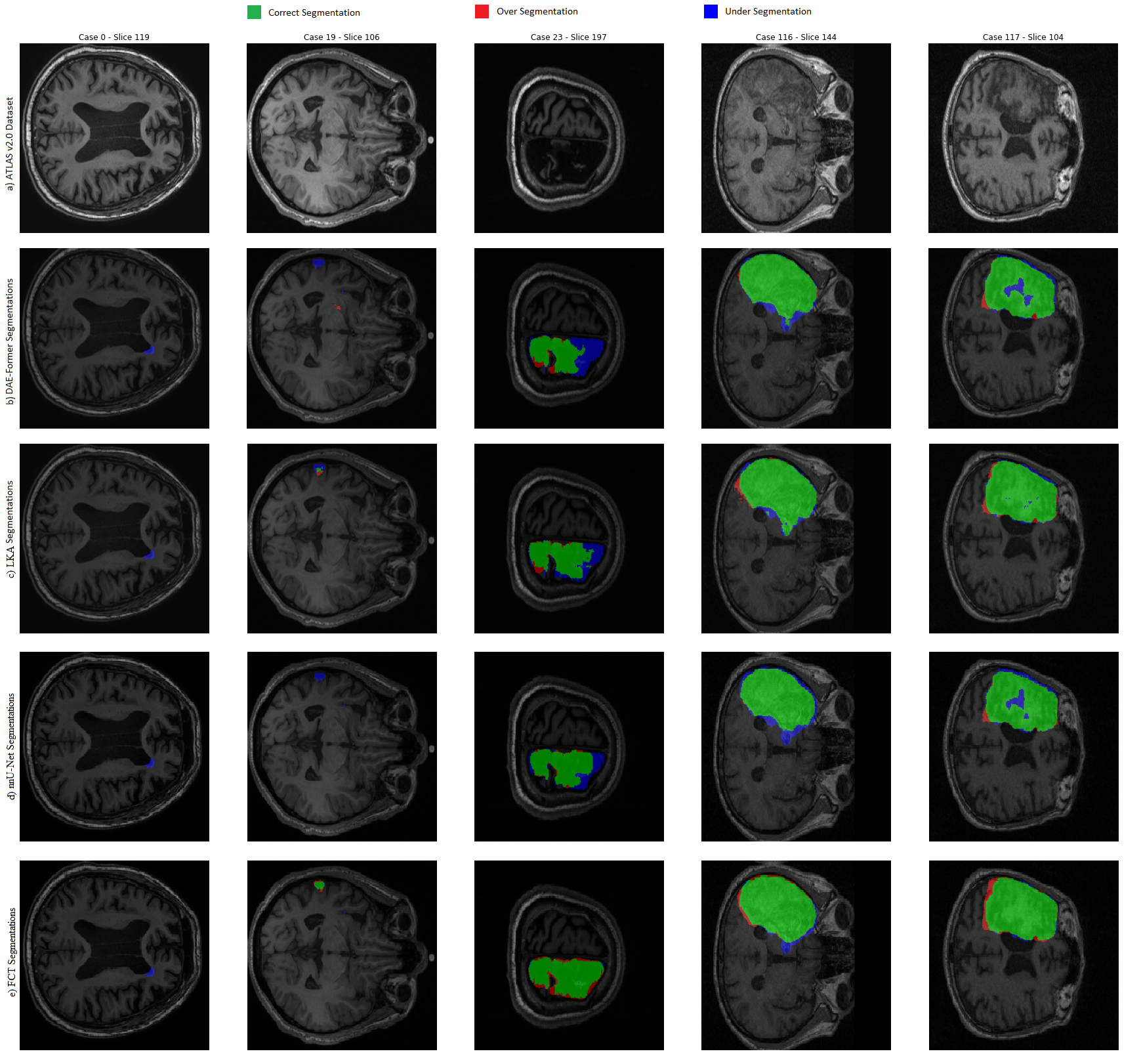}
  \caption{a) Sample of ATLAS v2.0 dataset,  b) Performance of the DAE-Former, \cite{azad2023dae},   c)  Performance of the LKA \cite{azad2024beyond},  d)  Performance of the D-LKA \cite{azad2024beyond},  e) Performance of the nnU-Net \cite{isensee2021nnu},  f)  Performance of the FCT \cite{tragakis2023fully}.}
  \label{fig:atlas}
\end{figure*}

Table \ref{tab:isles} and Table \ref{tab:atlas} summarize the performance of different models based on various ways of calculating the Dice score. Figures \ref{fig:isles} and \ref{fig:atlas} illustrate a qualitative analysis of the performance for different cases, highlighting various locations, sizes, and patterns of stroke infarcts.

\begin{table*}
  \centering
  \caption{Dice Scores for Different architecture using ISLES 2022 dataset; best results are shown in bold}
    \begin{tabular}{lcccc}
    \hline
    \textbf{Architectures} & \textbf{Aggregated Dice} & \multicolumn{2}{c}{\textbf{Slice-based Dice}} & \textbf{Volume-based Dice}  \\
    \cmidrule(lr){3-4} 
     & & \textbf{With Null Slices} & \textbf{Without Null Slices} &  \\
    \hline
    DAE-Former & 0.84284 & 0.67930 & 0.47472 & 0.68694  \\
    FCT & 0.84417 & 0.78966 & 0.55605 & 0.72042  \\ 
    LKA & 0.84824 & 0.71638 & 0.49948 & 0.70984  \\ 
    D-LKA & 0.87087 & 0.69694 & 0.50327 & 0.75644  \\ 
    nnU-Net & \bf{0.88649} & \bf{0.87679} & \bf{0.71166} & \bf{0.80044}  \\ 
    \hline
  \end{tabular}
  \label{tab:isles}
\end{table*}

\begin{table*}
  \centering
  \caption{Dice Scores for Different architecture using ATLAS v2.0 dataset; best results are shown in bold}
    \begin{tabular}{lcccc}
    \hline
    \textbf{Architectures} & \textbf{Aggregated Dice} & \multicolumn{2}{c}{\textbf{Slice-based Dice}} & \textbf{Volume-based Dice}  \\
    \cmidrule(lr){3-4} 
     & & \textbf{With Null Slices} & \textbf{Without Null Slices} &  \\
    \hline
    DAE-Former & 0.76849 & 0.73544 & 0.26141 & 0.39351  \\
    FCT & 0.80510 & 0.84771 & 0.37167 & 0.43506 \\ 
    LKA & 0.78569 & 0.76583 & 0.27653 & 0.40620  \\ 
    nnU-Net & \bf{0.83332} & \bf{0.90935} & \bf{0.50935} & \bf{0.57336}  \\ 
    \hline
  \end{tabular}
  \label{tab:atlas}
\end{table*}

Across both datasets, DAE-Former exhibited the weakest performance, while nnU-Net achieved the best results, although they were still not entirely favorable. In the comparison of Transformer-based architectures, FCT outperformed DAE-Former by a significant margin in most Dice scores, highlighting the importance of local information provided by CNNs for stroke segmentation.

While Transformer-based architectures have demonstrated success in various medical image segmentation tasks, they still struggle to achieve desirable performance for stroke segmentation. This challenge arises from the high variability exhibited in stroke infarcts across various features such as morphology, size, number, location, and pattern. To investigate the potential reasons for the performance gap between pure Transformer architectures and hybrid or CNN-based architectures, we aimed to conduct ablation studies to analyze the impact of data factors on the Transformer's learning process, which could potentially lead to weaker results.

\begin{figure}
  \centering
  \includegraphics[width=0.5\textwidth]{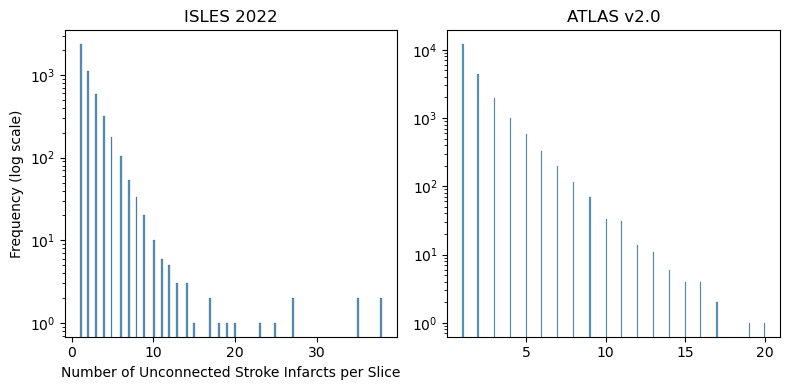}
  \caption{Distribution of the number of unconnected components for stroke infarcts per slice in each dataset.}
  \label{fig:histogram}
\end{figure}

Splitting the data based on infarct size across different slices posed a significant challenge and resulted in a reduction in the amount of available data. Therefore, we decided to exclude the study based on infarct size. Another important variability in stroke segmentation is the presence of multiple instances and the possibility of a large number of unconnected components within the same slice. To assess the robustness of DAE-Former to this variability, we conducted an ablation study, the results of which are summarized in Table \ref{tab:daeformer} for both the ISLES 2022 and ATLAS datasets. We excluded slices with more than 5 unconnected components from the training dataset of ISLES 2022 and slices with more than 3 unconnected components from the training dataset of ATLAS, while keeping the test dataset unchanged. These thresholds were chosen based on the histogram of the number of unconnected components in each dataset, as depicted in Figure \ref{fig:histogram}. As shown in the table, by excluding those slices, the DAE-Former network demonstrated improved performance across all Dice scores in both datasets, despite being trained with less data and evaluated on the same test dataset. The limited number of slices with a high number of unconnected components resulted in a weaker overall performance, indicating that Transformers are not robust to such variabilities with an imbalanced distribution of data.

\begin{table*}
  \centering
  \caption{An ablation study assessing the impact of the distribution of unconnected components in the training dataset on evaluation performance using the same dataset. In the selected training dataset, slices with a high number of unconnected components were excluded.}
    \begin{tabular}{lcccc}
    \hline
    \textbf{Dataset} & \textbf{Aggregated Dice} & \multicolumn{2}{c}{\textbf{Slice-based Dice}} & \textbf{Volume-based Dice}  \\
    \cmidrule(lr){3-4} 
     & & \textbf{With Null Slices} & \textbf{Without Null Slices} &  \\
    \hline
    ISLES 2022 (all train data) & 0.84284 & 0.67930 & 0.47472 & 0.68694  \\
    ISLES 2022 (selected train data) & 0.84656 & 0.69261 & 0.48199 & 0.70508 \\
    \hline
    ATLAS v2.0 (all train data) & 0.76849 & 0.73544 & 0.26141 & 0.39351  \\ 
    ATLAS v2.0 (selected train data) & 0.77815 & 0.77976 & 0.29254 & 0.40713  \\ 

    \hline
  \end{tabular}
  \label{tab:daeformer}
\end{table*}

Adapting CNNs to capture local information, particularly texture-related features, appears to be a more rational approach for brain stroke segmentation. When designing hybrid CNN-Transformer architectures, careful consideration of how to integrate Transformer-based and CNN-based blocks is essential. Furthermore, FCT achieved better results for most Dice scores compared to LKA and DLKA, suggesting that the global information captured by Transformer-based structures may be more beneficial than existing methods that modify CNN-based structures to extract global information for stroke segmentation.

The nnU-Net framework designed a simple U-shaped architecture with adjusted hyperparameters based on the data, as well as pre- and post-processing steps that can enhance results. The CNN blocks used in these architectures have a size of 3, primarily providing local information rather than global information. The superior results obtained using the nnU-Net framework compared to other architectures highlight the importance of preprocessing and postprocessing steps for stroke segmentation. The complex characteristics of stroke segmentation require careful consideration when designing automatic pipelines. It's worth noting that the results for stroke segmentation are still not as advanced as those for other tasks such as brain tumor segmentation \cite{ranjbarzadeh2023brain}, indicating the need for further refinements and considerations in designing automated pipelines to make them applicable in clinical settings.

In particular, the performance of all methods for all architectures was higher for the ISLES 2022 dataset compared to the ATLAS dataset. This disparity may be attributed to the imaging modality used in ATLAS (T1-weighted), where stroke infarcts are less visible compared to the DWI modality. Additionally, the imbalance issue in the ATLAS dataset was more severe than in ISLES. As evident from the figures, all methods exhibited better performance for large infarcts compared to small ones. For tiny infarcts where only a few pixels were labeled as stroke in the entire slice, most architectures struggled to segment these infarcts accurately.

Furthermore, the Aggregated Dice metric shows less variation in performance across different architectures for this task, likely due to its bias towards larger sizes in tasks containing variable sizes for segmentation such as stroke segmentation. When considering null slices in calculating the Dice score for this task, wherein many slices may lack infarction, it can lead to an illusion of higher performance. However, this issue is mitigated when using slice-based Dice score calculation, which excludes null slices, as well as volume-based Dice score calculation.

\section{Conclusion}

In this study, we evaluated different deep architectures for brain stroke segmentation in MRI images. We selected four types of networks for this evaluation: a pure Vision Transformer-based architecture (DAE-Former), one advanced fully convolutional Transformer (FCT), two CNN-based networks with visual attention mechanisms in their design (LKA, D-LKA), and the widely-used self-adaptive nnU-Net framework. All these networks have shown superior performance in various segmentation tasks and datasets. We assessed their performance on two publicly available datasets: ISLES 2022 and ATLAS v2.0.

Based on the results, nnU-Net outperformed other baseline methods by a large margin, despite having the simplest design, while the pure Transformer network (DAE-Former) exhibited the weakest performance. This suggests that the incorporation of CNN layers in the proposed architectures for brain stroke segmentation provides more informative features compared to the information that Transformers can provide alone. Our evaluations indicate that local information is more effective than global information for stroke segmentation, likely due to the high variability in stroke shapes, sizes, locations, and patterns that global information struggles to adequately capture for segmentation purposes.

We also conducted an ablation study to explore the impact of the number of unconnected components in the training dataset on the performance of the pure Transformer architecture. This investigation aimed to understand why Transformers exhibit weaker performance in stroke segmentation compared to other tasks where this issue is not observed. Our study revealed that limiting the distribution of the number of unconnected infarcts in each slice, and excluding slices with a high number of unconnected components, led to improved performance on the same test dataset. This suggests that Transformers are not robust to such features, where there is an imbalanced distribution across the dataset, with limited slices representing certain parts of the distribution. Further analysis could involve examining the robustness of Transformer-based architectures based on advanced morphological features, particularly for tasks like stroke segmentation that exhibit high variability.

While our research revealed that the simplest architectural design achieved the highest performance, it was largely attributable to the preprocessing and post-processing steps incorporated into nnU-Net. Exploring more advanced architectures, coupled with appropriate data processing techniques, has the potential to significantly enhance results, given that current outcomes are not satisfactory. For future endeavors, researchers could focus on carefully selecting network elements based on the characteristics of stroke segmentation to develop a potentially applicable network.

Although we carefully selected different architectural designs to examine their performance in stroke segmentation, it is important to acknowledge the limitations of our study. Firstly, there are numerous architectures proposed for medical image segmentation in recent years, and we only considered some of them. The results may not be generalizable to other architectures, whether they are Transformer-based, hybrid, or CNN-based. Secondly, due to the time required to train baseline models, we were unable to perform an extensive grid search for tuning the hyperparameters. Therefore, we conducted the grid search within a close range of the originally proposed hyperparameters. This could affect the results, and the architectures may perform differently in other configurations.

\section{Acknowledgment}
The findings presented in this paper have emerged from a project funded by the Qatar Japan Research Collaboration Research Program under grant number M-QJRC-2023-313. The authors extend their sincere gratitude to Marubeni and Qatar University for their consistent and generous support.

\bibliographystyle{unsrt}
\bibliography{refs}

\end{document}